\documentclass{amsart}
\usepackage{graphicx}
\usepackage[round,colon]{natbib}

\newcommand{\calH}{H}
\newcommand{\calR}{R}
\newcommand{\cal}{\mathcal}

\newtheorem{theorem}{Theorem}
\newtheorem{lemma}{Lemma}

\begin{document}

\title[Confidence measures via multiscale bootstrap]{Frequentist and
Bayesian measures of confidence via multiscale bootstrap for testing
three regions}

\author{Hidetoshi Shimodaira}
\address{Department of Mathematical and Computing Sciences\\
Tokyo Institute of Technology\\
2-12-1 Ookayama, Meguro-ku, Tokyo 152-8552, Japan}
\email{shimo@is.titech.ac.jp} 



\date{July 1, 2008}

\begin{abstract}

A new computation method of frequentist $p$-values and Bayesian
posterior probabilities based on the bootstrap probability is discussed
for the multivariate normal model with unknown expectation parameter
vector.  The null hypothesis is represented as an arbitrary-shaped
region.  We introduce new parametric models for the scaling-law of
bootstrap probability so that the multiscale bootstrap method, which was
designed for one-sided test, can also computes confidence measures of
two-sided test, extending applicability to a wider class of hypotheses.
Parameter estimation is improved by the two-step multiscale bootstrap
and also by including higher-order terms.  Model selection is important
not only as a motivating application of our method, but also as an
essential ingredient in the method.  A compromise between frequentist
and Bayesian is attempted by showing that the Bayesian posterior
probability with an noninformative prior is interpreted as a frequentist
$p$-value of ``zero-sided'' test.
\end{abstract}

\maketitle

\section{Introduction} 
\label{sec:intro}

Let $Y=(Y_1,\ldots,Y_{m+1})$ be a random vector of dimension $m+1$ for
some integer $m\ge1$, and $y=(y_1,\ldots,y_{m+1})\in\mathbb{R}^{m+1}$
be its observed value. Our argument is based on the multivariate normal
model with unknown mean vector
$\mu=(\mu_1,\ldots,\mu_{m+1})\in\mathbb{R}^{m+1}$ and covariance
identity $I_{m+1}$,
\begin{equation} \label{eq:ynorm}
 Y \sim N_{m+1}(\mu,I_{m+1}),
\end{equation}
where the probability with respect to (\ref{eq:ynorm}) will be denoted
as $P(\cdot|\mu)$.  Let ${\calH}_0\subset\mathbb{R}^{m+1}$ be an
arbitrary-shaped region. The subject of this paper is to compute
measures of confidence for testing the null hypothesis $\mu\in{\calH}_0$. Observing $y$, we compute a frequentist $p$-value, denoted
$p({\calH}_0|y)$, and also a Bayesian posterior probability $\pi({\calH}_0 | y)$ with a noninformative prior density $\pi(\mu)$ of $\mu$.

This is the {\em problem of regions} discussed in literature;
\cite{bib:Efro:Hall:Holm:96:BCL}, \cite{bib:Efro:Tibs:98:PR}, and
\cite{bib:Shim:2002:AUT, bib:Shim:2004:AUT, bib:Shimo:2008:TRN}.  The
confidence measures were calculated by the bootstrap methods for
complicated application problems such as the variable selection of
regression analysis and phylogenetic tree selection of molecular
evolution.  These model selection problems are motivating applications
for the issues discussed in this paper, and the normal model of
(\ref{eq:ynorm}) is a simplification of reality. Let ${\cal
X}=\{x_1,\ldots,x_n\}$ be a sample of size $n$ in application
problems. We assume there exists a transformation, depending on $n$,
from ${\cal X}$ to $y$ so that $Y$ is approximately normalized. We
assume only the existence of such a transformation, and do not have to
consider its details.  Since we work only on the transformed variable
$Y$ in this paper for developing the theory, readers may refer to the
literature above for the examples of applications.  Before the problem
formulation is given in Section~\ref{sec:formulation}, our methodology
is illustrated in simple examples below in this section.

The simplest example of ${\calH}_0$ would be the half space of
$\mathbb{R}^{m+1}$,
\begin{equation} \label{eq:h0s2}
 {\calH}_0' : \mu_{m+1}\le 0,
\end{equation}
where the notation ${\calH}_0'$, instead of ${\calH}_0$, is used to
distinguish this case from another example given in (\ref{eq:h0s3}).
Only $\mu_{m+1}$ is involved in this ${\calH}_0'$, and one-dimensional
normal model $Y_{m+1} \sim N(\mu_{m+1},1)$ is considered.  Taking
$\mu_{m+1}>0$ as an alternative hypothesis and denoting the cumulative
distribution function of the standard normal as $\Phi(\cdot)$ with
density $\phi(\cdot)$, the unbiased frequentist $p$-value is given as
$p({\calH}_0' | y)= \Phi(-y_{m+1})$.

A slightly complex example of ${\calH}_0$ is
\begin{equation} \label{eq:h0s3}
 {\calH}_0 : -d \le \mu_{m+1}\le 0
\end{equation}
for $d>0$. The rejection regions are $y_{m+1}>c$ and $y_{m+1}< -d -c$
with a critical constant $c$, which is obtained as a solution of the
equation
\begin{equation} \label{eq:defc}
 \Phi(-c) + \Phi(-d-c) =\alpha 
\end{equation}
for a specified significance level $0<\alpha<1$.  The left hand side of
 (\ref{eq:defc}) is the rejection probability $P(Y_{m+1}>c \vee
 Y_{m+1}<-d-c | \mu)$ when $\mu$ is on the boundary of ${\calH}_0$,
 i.e., $\mu_{m+1}=0$ or $\mu_{m+1}=-d$. The frequentist $p$-value is
 defined as the infimum of $\alpha$ such that ${\calH}_0$ can be
 rejected. This becomes $p({\calH}_0|y)=\Phi(-y_{m+1}) +
 \Phi(-d-y_{m+1})$ for $y_{m+1}\ge -d/2$ and
 $p({\calH}_0|y)=\Phi(y_{m+1}) + \Phi(d+y_{m+1})$ for $y_{m+1}\le
 -d/2$. Considering the case, say,
\begin{equation} \label{eq:numex1}
d=1,\quad y_{m+1}=-0.1,
\end{equation}
we obtain $p({\calH}_0' | y)=0.540$ and $p({\calH}_0 | y)=0.724$.

These two simple cases of ${\calH}_0$ and ${\calH}_0'$ exhibit what
\cite{bib:Efro:Tibs:98:PR} called {\em paradox} of frequentist
$p$-values. Our simple examples of (\ref{eq:h0s2}) and (\ref{eq:h0s3})
suffice for this purpose, although they had actually used the spherical
shell example explained later in Section~\ref{sec:three}.
\cite{bib:Efro:Tibs:98:PR} indicated that a confidence measure should be
monotonically increasing in the order of set inclusion of the
hypothesis. Noting ${\calH}_0 \subset {\calH}_0'$, therefore, it should
be $p({\calH}_0|y)\le p({\calH}_0'|y)$, but it is not.  This kind of
``paradox'' cannot occur with Bayesian methods, and $\pi({\calH}_0|y)\le
\pi({\calH}_0'|y)$ holds always. Considering the flat prior $\pi(\mu)=$
const, say, the posterior distribution of $\mu$ given $y$ becomes
\begin{equation} \label{eq:postmuflat}
 \mu|y \sim N_{m+1}(y,I_{m+1}),
\end{equation}
and the posterior probabilities for the case (\ref{eq:numex1}) are
$\pi({\calH}_0' | y) = \Phi(-y_{m+1}) = 0.540$ and $\pi({\calH}_0 | y) =
\Phi(-y_{m+1}) - \Phi(-d-y_{m+1}) = 0.356$.  The ``paradox'' of
frequentist $p$-values may be nothing surprise for a frequentist
statistician, but a natural consequence of the fact that
$p({\calH}_0'|y)$ is for a one-sided test and $p({\calH}_0|y)$ is for a
two-sided test; The power of testing is higher, i.e., $p$-values are
smaller, for an appropriately formulated one-sided test than a two-sided
test.  In this paper, we do not intend to argue the philosophical
question of whether to be frequentist or to be Bayesian, but discuss
only computation of these two confidence measures.

Computation of the confidence measures is made by the bootstrap
resampling of \cite{bib:Efro:79:BMA}. Let ${\cal
X}^*=\{x_1^*,\ldots,x_{n'}^*\}$ be a bootstrap sample of size $n'$
obtained by resampling with replacement from ${\cal X}$. The idea of
bootstrap probability, which is introduced first by
\citet{bib:Fels:85:CLP} to phylogenetic inference, is to generate ${\cal
X}^*$ many times, say $B$, and count the frequency $C$ that a hypothesis
of interest is supported by the bootstrap samples. The bootstrap
probability is computed as $C/B$.  Recalling the transformation to get
$y$ from ${\cal X}$, we get $Y^*$ by applying the same transformation to
${\cal X}^*$. For typical problems, the variance of $Y^*$ is
approximately proportional to the factor
\[
 \sigma^2=\frac{n}{n'}
\]
as mentioned in \cite{bib:Shimo:2008:TRN}.  Although we generate ${\cal
X}^*$ in practice, we only work on $Y^*$ in this paper. More
specifically, we formally consider the parametric bootstrap
\begin{equation} \label{eq:yboot}
 Y^* | y \sim N_{m+1}(y,\sigma^2 I_{m+1}),
\end{equation}
which is analogous to (\ref{eq:ynorm}) but the scale $\sigma$ is
introduced for multiscale bootstrap.  The bootstrap probability is
defined as
\begin{equation} \label{eq:bp}
 \alpha_{\sigma^2}({\calH}_0 | y)
=P_{\sigma^2}(Y^* \in {\calH}_0 | y),
\end{equation}
where $P_{\sigma^2}(\cdot|y)$ denotes the probability with respect to
(\ref{eq:yboot}). For computing a crude confidence measure, we set
$\sigma=1$, or $n'=n$ in terms of ${\cal X}^*$, so that the distribution
(\ref{eq:yboot}) for $Y^*$ is equivalent to the posterior
(\ref{eq:postmuflat}) for $\mu$. This gives an interpretation of the
bootstrap probability that $\alpha_{1}({\calH}_0|y)=\pi({\calH}_0|y)$
for any ${\calH}_0$ under the flat prior.  In the multiscale bootstrap
of \cite{bib:Shim:2002:AUT, bib:Shim:2004:AUT, bib:Shimo:2008:TRN},
however, we may intentionally alter the scale from $\sigma=1$, or to
change $n'$ from $n$ in terms of ${\cal X}^*$ for computing
$p({\calH}_0|y)$.  Let $\sigma_1,\ldots,\sigma_M$ be $M$ different
values of scale, which we specify in advance.  In our numerical
examples, $M=13$ scales are equally spaced in log-scale between
$\sigma_1=1/3$ and $\sigma_{13}=3$.  For each $i=1,\ldots,M$, we
generate ${\cal X}^*$ with scale $\sigma_i$ for $B_i$ times, and observe
the frequency $C_i$.  The observed bootstrap probability is $\hat
\alpha_{\sigma^2_i}= C_i/B_i$.

How can we use the observed $\hat\alpha_{\sigma^2_1},\ldots,
\hat\alpha_{\sigma^2_M}$ for computing $p({\calH}_0|y)$?  Let us assume
that ${\calH}_0$ can be expressed as ({\ref{eq:h0s3}}) but we are unable
to observe the values of $y_{m+1}$ and $d$.  Nevertheless, by fitting
the model $ \alpha_{\sigma^2}({\calH}_0|y) = \Phi(-y_{m+1}/\sigma)
-\Phi(-(d+y_{m+1})/\sigma)$ to the observed
$\hat\alpha_{\sigma^2_1},\ldots, \hat\alpha_{\sigma^2_M}$, we may
compute an estimate $\hat\varphi$ of the parameter vector
$\varphi=(y_{m+1},d)$ with constraints $d>0$ and $y_{m+1}>-d/2$. The
confidence measures are then computed as $p({\calH}_0|y)=\Phi(-\hat
y_{m+1}) + \Phi(-\hat d-\hat y_{m+1})$ and $\pi({\calH}_0 | y) =
\Phi(-\hat y_{m+1}) - \Phi(-\hat d-\hat y_{m+1})$. In case we are not
sure which of (\ref{eq:h0s2}) and (\ref{eq:h0s3}) is the reality, we may
also fit $ \alpha_{\sigma^2}({\calH}_0'|y) = \Phi(-y_{m+1}/\sigma)$ to
the observed $\hat\alpha_{\sigma^2_i}$'s and compare the AIC values
\citep{bib:Akai:74:NLS} for model selection. In practice, we prepare
collection of such models describing the scaling-law of bootstrap
probability, and choose the model which minimizes the AIC value.

\section{Formulation of the problem} \label{sec:formulation}

\begin{figure}
\includegraphics[width=0.75\textwidth]{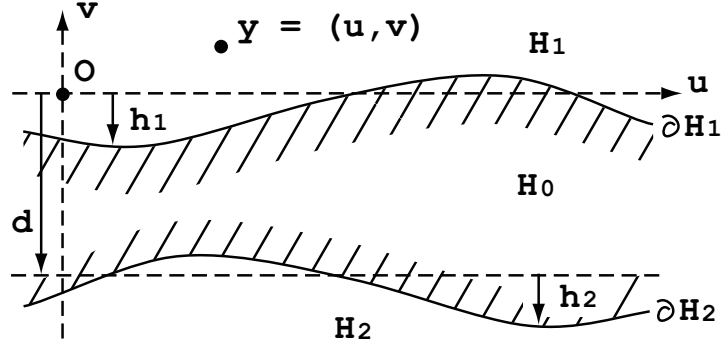}
\caption{Region ${\calH}_0 \subset \mathbb{R}^{m+1}$ is the shaded area
between surfaces $\partial {\calH}_1$ and $\partial {\calH}_2$.}
\label{fig:h03}
\end{figure}

The examples in Section~\ref{sec:intro} were very simple because the
boundary surfaces of the regions are flat. In the following sections, we
work on generalizations of (\ref{eq:h0s2}) and (\ref{eq:h0s3}) by
allowing curved boundary surfaces.  For convenience, we denote $y=(u,v)$
with $u=(y_1,\ldots,y_m)$ and $v=y_{m+1}$. Similarly, we denote
$\mu=(\theta,\mu_{m+1})$ with $\theta=(\mu_1,\ldots,\mu_m)\in
\mathbb{R}^m$.  As shown in Fig.~\ref{fig:h03}, we consider the region
of the form $ {\calH}_0 = \{ (\theta,\mu_{m+1}) \mid -d - h_2(\theta)
\le \mu_{m+1} \le -h_1(\theta),\, \theta\in \mathbb{R}^m \}$, where
$h_1(\theta)$ and $h_2(\theta)$ are arbitrary functions of $\theta$.
This region will reduce to (\ref{eq:h0s3}) if $h_1(\theta)= h_2(\theta)=0$
for all $\theta$.  The region may be abbreviated as
\begin{equation} \label{eq:h03}
{\calH}_0: -d - h_2(\theta) \le
\mu_{m+1} \le -h_1(\theta).
\end{equation} 
 Two other regions ${\calH}_1: \mu_{m+1} \ge -h_1(\theta)$ and ${\calH}_2: \mu_{m+1} \le -d-h_2(\theta)$ as well as two boundary surfaces
$\partial {\calH}_1: \mu_{m+1}=-h_1(\theta)$ and $\partial {\calH}_2:
\mu_{m+1}=-d-h_2(\theta)$ are also shown in Fig.~\ref{fig:h03}.  
We define  ${\calH}_0'={\calH}_0 \cup {\calH}_2$, or equivalently as
\begin{equation} \label{eq:h02}
 {\calH}_0': \mu_{m+1} \le -h_1(\theta).
\end{equation}
The boundary surfaces of the hypotheses are $\partial {\calH}_0 =
\partial {\calH}_1 \cup \partial {\calH}_2$ for the region ${\calH}_0$,
and $\partial {\calH}'_0 = \partial {\calH}_1$ for the region
${\calH}'_0$.

We do not have to specify the functional forms of $h_1$ and $h_2$ for our
theory, but assume that the magnitude of $h_1$ and $h_2$ is very small.
Technically speaking, $h_1$ and $h_2$ are {\em nearly flat} in the sense
of \cite{bib:Shimo:2008:TRN}. Introducing an artificial parameter
$\lambda$, a function $h$ is called nearly flat when
$\sup_{\theta\in\mathbb{R}^m} |h(\theta)| = O(\lambda)$ and $L^1$-norms
of $h$ and its Fourier transform are bounded. We develop asymptotic
theory as $\lambda\to0$, which is analogous to $n\to\infty$ with the
relation $\lambda=1/\!\sqrt{n}$.

The whole parameter space is partitioned into two regions as ${\calH}_0'
\cup {\calH}_1 = \mathbb{R}^{m+1}$ or three regions as ${\calH}_0 \cup
{\calH}_1 \cup {\calH}_2 = \mathbb{R}^{m+1}$. These partitions are
treated as disjoint in this paper by ignoring measure-zero sets such as
${\calH}_0' \cap {\calH}_1 = \partial {\calH}_1$.  Bootstrap methods for
computing frequentist confidence measures are well developed in the
literature as reviewed in Section~\ref{sec:two}. The main contribution
of our paper is then given in Section~\ref{sec:three} for the case of
three regions.  In Section~\ref{sec:bayes}, this new computation method
is used also for Bayesian measures of \cite{bib:Efro:Tibs:98:PR}.  Note
that the flat prior $\pi(\mu)=$ const in the previous section was in
fact carefully chosen so that $\pi({\calH}_0'|y)=p({\calH}_0'|y)$ for
(\ref{eq:h0s2}).  This same $\pi(\mu)$ led to $\pi({\calH}_0|y)\neq
p({\calH}_0|y)$ for (\ref{eq:h0s3}).  Our definition of ${\calH}_0$
given in (\ref{eq:h03}) is a simplest formulation, yet with a reasonable
generality for applications, to observe such an interesting difference
between the two confidence measures.

Multiscale bootstrap computation of the confidence measures for the
three regions case is described in Section~\ref{sec:model}. Simulation
study and some discussions are given in Section~\ref{sec:simulation} and
\ref{sec:discussion}, respectively. Mathematical proofs are mostly given
in Appendix.

\section{Frequentist measures of confidence for testing two regions}
\label{sec:two}

In this section, we review the multiscale bootstrap of
\cite{bib:Shimo:2008:TRN} for computing a frequentist $p$-value of
``one-sided'' test of ${\calH}_0'$.  Let $z=-\Phi^{-1}(\alpha)$ be the
inverse function of $\alpha=\Phi(-z)$. The bootstrap $z$-value of
${\calH}_0'$, defined as
$z_{\sigma^2}({\calH}_0'|y)=-\Phi^{-1}(\alpha_{\sigma^2}({\calH}_0'|y))$,
is convenient to work with.  By multiplying $\sigma$ to it, $\sigma
z_{\sigma^2}({\calH}_0'|y)$ is called the normalized bootstrap
$z$-value.  Theorem~1 of \cite{bib:Shimo:2008:TRN}, as reproduced below,
states that the $z$-value of $p({\calH}_0'|y)$ is obtained by
extrapolating the normalized bootstrap $z$-value to $\sigma^2=-1$, or
equivalently $n'=-n$ in terms of ${\cal X}^*$.
\begin{theorem} \label{thm:onesided} Let ${\calH}_0'$ be a region of
 (\ref{eq:h02}) with nearly flat $h_1$. Given ${\calH}_0'$ and $y$,
 consider the normalized bootstrap $z$-value as a function of
 $\sigma^2$; We denote it by $\psi(\sigma^2)= - \sigma
 \Phi^{-1}(\alpha_{\sigma^2}({\calH}_0' | y))$. Let us define a
 frequentist $p$-value as
\begin{equation} \label{eq:ph02}
 p({\calH}_0' | y) = \Phi(-\psi(-1)),
\end{equation}
and assume that the right hand side exists.  Then for $\mu \in
\partial{\calH}_0'$ and $0<\alpha<1$,
\begin{equation} \label{eq:eh02}
 P(p({\calH}_0' |Y)<\alpha | \mu)= \alpha + O(\lambda^3),
\end{equation}
meaning that the coverage error, i.e., the difference between the
 rejection probability and $\alpha$, vanishes asymptotically as
 $\lambda\to 0$, and that the $p$-value, or the associated hypothesis
 testing, is ``similar on the boundary'' asymptotically.
\end{theorem}

\paragraph{Proof} Here we show only an outline of the proof by allowing
 the coverage error of $O(\lambda^2)$, instead of $O(\lambda^3)$, in
 (\ref{eq:eh02}). This is a brief summary of the argument given in
 \cite{bib:Shimo:2008:TRN}. First define the expectation operator
 ${\cal E}_{\sigma^2}$ for a  nearly flat function $h$ as
\[
 ({\cal E}_{\sigma^2} h)(u) := E_{\sigma^2} ( h(U^*)| u ),
\]
where $E_{\sigma^2}(\cdot)$ on the right hand side
denotes the expectation with respect to (\ref{eq:yboot}), that is, for
$Y^*=(U^*,V^*)$ with
\[
 U^*|u \sim N_m(u,\sigma^2 I_m),\quad
 V^*|v \sim N(v,\sigma^2).
\]
Using the expectation operator, we next define two quantities
\[
 z_1 = -\frac{v+{\cal E}_{\sigma^2} h_1(u)}{\sigma},\quad 
 \epsilon_1 = -\frac{h_1(U^*) - {\cal E}_{\sigma^2} h_1(u)}{\sigma},
\]
and work on the bootstrap probability as
\begin{eqnarray}
 \alpha_{\sigma^2}({\calH}_0'|y) &=&
P_{\sigma^2}(V^* \le -h_1(U^*)|y) \nonumber\\
&=& E_{\sigma^2}\left( \Phi(z_1 + \epsilon_1) | u \right)\nonumber\\
&=& E_{\sigma^2}\left( \Phi(z_1)+ \phi(z_1) \epsilon_1  | u \right) +
 O(\lambda^2)\nonumber\\
&=& \Phi(z_1)+ O(\lambda^2). \label{eq:bph02}
\end{eqnarray}
The third equation is obtained by the Taylor series around $z_1$,
and the last equation is obtained by $E_{\sigma^2}(\epsilon_1|u)=0$.
Rearranging (\ref{eq:bph02}), we then get the scaling-law of the
normalized bootstrap $z$-value as
\begin{equation} \label{eq:scalelawh02}
 \psi(\sigma^2) = v + {\cal E}_{\sigma^2} h_1(u) + O(\lambda^2).
\end{equation}
On the other hand, eq.~(5.10) of \cite{bib:Shimo:2008:TRN} shows, by
utilizing Fourier transforms of surfaces, that (\ref{eq:eh02})
holds with coverage error $O(\lambda^2)$ for a $p$-value defined as
\begin{equation} \label{eq:phinvh02}
 p({\calH}_0'|y) = \Phi(-v-{\cal E}_{-1}h_1(u)) + O(\lambda^2).
\end{equation}
The proof completes by combining (\ref{eq:scalelawh02}) and
(\ref{eq:phinvh02}).
\qed

A hypothesis testing is to reject ${\calH}_0'$ when observing
$p({\calH}_0' |y)<\alpha$ for a specified significance level, say,
$\alpha=0.05$, and otherwise not to reject ${\calH}_0'$. The left hand
side of (\ref{eq:eh02}) is the rejection probability, which should be
$\le\alpha$ for $\mu\in {\calH}'_0$ and $\ge\alpha$ for $\mu\not\in
{\calH}'_0$ to claim the unbiasedness of the test.  On the other hand,
the test is called similar on the boundary when the rejection
probability is equal to $\alpha$ for $\mu\in \partial{\calH}_0'$.  In
this paper, we implicitly assume that $p({\calH}_0'|y)$ is decreasing as
$y$ moves away from ${\calH}_0'$. The rejection probability increases
continuously as $\mu$ moves away from ${\calH}_0'$. This assumption is
justified when $\lambda$ is sufficiently small so that the behavior of
$p({\calH}_0'|y)$ is not very different from that for (\ref{eq:h0s2}).
Therefore, (\ref{eq:eh02}) implies that the $p$-value is approximately
unbiased asymptotically as $\lambda\to0$.

We can think of a procedure for calculating $p({\calH}_0' | y)$ based on
(\ref{eq:ph02}).  In the procedure, the functional form of
$\psi(\sigma^2)$ should be estimated from the observed
$\hat\alpha_{\sigma^2_i}$'s using parametric models.  Then an
approximately unbiased $p$-value is computed by extrapolating
$\psi(\sigma^2)$ to $\sigma^2=-1$.  Our procedure works fine for the
particular ${\calH}_0'$ of (\ref{eq:h0s2}), because
$\psi(\sigma^2)=y_{m+1}$ and $p({\calH}_0' | y)=\Phi(-y_{m+1}) =
\Phi(-\psi(-1))$.  Our procedure works fine also for any ${\calH}_0'$ of
(\ref{eq:h02}) when the boundary surface $\partial {\calH}_0'$ is
smooth. The model is given as $\psi(\sigma^2)=\beta_0 + \beta_1 \sigma^2
+ \beta_2 \sigma^4 + \beta_3 \sigma^6 + \cdots$ using parameters
$\varphi=(\beta_0,\beta_1,\ldots)$, and thus an approximately unbiased
$p$-value can be computed by $p({\calH}_0'|y)=\Phi( -\hat\beta_0 +
\hat\beta_1 - \hat\beta_2 + \hat\beta_3 -\cdots)$.  It may be
interesting to know that the parameters are interpreted as geometric
quantities; $\beta_0$ is the distance from $y$ to the surface $\partial
{\calH}_0'$, $\beta_1$ is the mean curvature of the surface, and
$\beta_j$, $j\ge2$, is related to $2j$-th derivatives of $h_1$.

However, the series expansion above does not converge, i.e., $\psi(-1)$
does not exist, when $\partial {\calH}_0'$ is nonsmooth.  For example,
$\psi(\sigma^2)=\beta_0 + \beta_1 \sqrt{\sigma^2}$ serves as a good
approximating model for cone-shaped ${\calH}_0'$, for which $\psi(-1)$
does not take a value of $\mathbb{R}$.  This observation agrees with the
fact that an unbiased test does not exist for cone-shaped ${\calH}_0'$
as indicated in the argument of \cite{bib:Lehm:52:TMH}.  Instead of
(\ref{eq:ph02}), the modified procedure of \cite{bib:Shimo:2008:TRN}
calculates a $p$-value defined as
\begin{equation} \label{eq:pk}
 p_k({\calH}_0'|y) = \Phi\left\{
-\sum_{j=0}^{k-1} \frac{(-1-\sigma_0^2)^j}{j!} 
\frac{\partial^j \psi(\sigma^2 )}
{\partial (\sigma^2)^j}\Bigr|_{\sigma_0^2} \right\}
\end{equation}
 for an integer $k>0$ and a real number $\sigma^2_0>0$. This is to
extrapolate $\psi(\sigma^2)$ back to $\sigma^2=-1$ by using the first
$k$ terms of the Taylor series around $\sigma_0^2$.  The coverage error
in (\ref{eq:eh02}) should reduce as $k$ increases, but then the
rejection region violates the desired property called monotonicity in
the sense of \citet{bib:Lehm:52:TMH} and
\cite{bib:Perl:Wu:1999:ENT,bib:Perl:Wu:2003:OVL}. For taking the
balance, we chose $k=3$ and $\sigma_0^2=1$ for numerical examples in
this paper.

\section{Frequentist measures of confidence for testing three regions}
\label{sec:three}

The following theorem is our main result for computing a frequentist
$p$-value of ``two-sided'' test of ${\calH}_0$. The proof is given in
Appendix~\ref{app:proof-twosided}.
\begin{theorem} \label{thm:twosided}
Let ${\calH}_0$ be a region of (\ref{eq:h03}) with nearly flat $h_1$
and $h_2$. Given ${\calH}_0$ and $y$, consider the approximately
unbiased $p$-value $p({\calH}_i|y)$ by applying
Theorem~\ref{thm:onesided} to ${\calH}_i$ for $i=1,2$. Assuming these
two $p$-values exist, let us define a frequentist $p$-value of ${\calH}_0$ as
\begin{equation} \label{eq:ph03}
 p({\calH}_0 | y) = 1 - | p({\calH}_1|y) - p({\calH}_2|y) |.
\end{equation}
For example, (\ref{eq:ph03}) holds for the exact $p$-value of
 (\ref{eq:h0s3}) defined in Section~\ref{sec:intro}.  Then for $\mu\in
 \partial {\calH}_0 = \partial {\calH}_1 \cup \partial {\calH}_2$ and
 $0<\alpha<1$,
\begin{equation}  \label{eq:eh03}
 P(p({\calH}_0 |Y)<\alpha | \mu)= \alpha + O(\lambda^2),
\end{equation}
meaning that $p({\calH}_0|y)$ is approximately unbiased
 asymptotically as $\lambda\to 0$.
\end{theorem}

\begin{figure}
\includegraphics[width=0.70\textwidth]{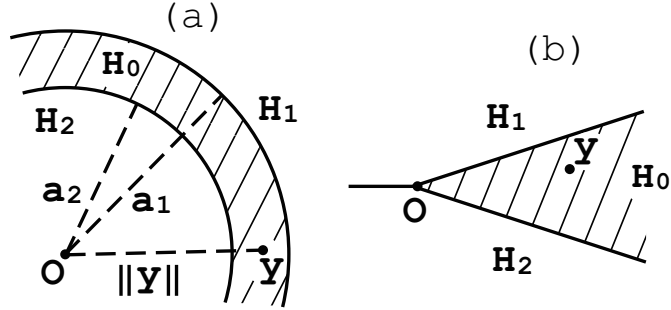}
\caption{(a)~Spherical shell region. (b)~Cone-shaped region
(Section~\ref{sec:simulation}).}  \label{fig:exregions}
\end{figure}

For illustrating the methodology, let us work on the spherical shell
example of \cite{bib:Efro:Tibs:98:PR}, for which we can still compute
the exact $p$-values to verify our methods. The region of interest is
${\calH}_0 : a_2 \le \|\mu\| \le a_1$ as shown in Panel~(a) of
Fig.~\ref{fig:exregions}. We consider the case, say,
\[
 m+1=4,\quad a_1=6,\quad a_2=5,\quad \|y\|=5.9,
\]
so that this region is analogous to (\ref{eq:numex1}) except for the
curvature.  The exact $p$-value for ${\calH}_1 : \|\mu\| \ge a_1$ is
easily calculated knowing that $\|Y\|^2$ is distributed as the
chi-square distribution with degrees of freedom $m+1$ and noncentrality
$\|\mu\|^2$. Writing this random variable as $\chi^2_{m+1}(\|\mu\|^2)$,
the exact $p$-value is $p({\calH}_1 | y) = P(\chi^2_{m+1}(a_1^2) \le
\|y\|^2)=0.362$, that is, the probability of observing $\|Y\|\le\|y\|$
for $\|\mu\|=a_1$.  Similarly, the exact $p$-value for ${\calH}_2 :
\|\mu\| \le a_2$ is $p({\calH}_2 | y) = P(\chi^2_{m+1}(a_2^2) \ge
\|y\|^2)=0.267$.  In a similar way as for (\ref{eq:h0s3}), the exact
$p$-value for ${\calH}_0$ is computed numerically as $p({\calH}_0 | y) =
0.907$, although the procedure is a bit complicated as explained below.
We first consider the critical constants $c_1$ and $c_2$ for the
rejection regions ${\calR}_1 = \{ y \mid \|y\| < a_1 - c_1\}$ and
${\calR}_2 = \{ y \mid \|y\| > a_2 + c_2\}$. By equating the rejection
probability to $\alpha$ for $\mu\in \partial {\calH}_0$, that is, $
P(\chi^2_{m+1}(a_i^2) < (a_1-c_1)^2) +
P(\chi^2_{m+1}(a_i^2)>(a_2+c_2)^2) = \alpha$ for $i=1,2$, we may get the
solution numerically as $c_1=1.331$ and $c_2=1.903$ for $\alpha=0.05$,
say. The $p$-value is defined as the infimum of $\alpha$ such that
${\calH}_0$ can be rejected.

To check if Theorem~\ref{thm:twosided} is ever usable, we first compute
(\ref{eq:ph03}) using the exact values of $p({\calH}_1|y)$ and
$p({\calH}_2|y)$. Then we get $p({\calH}_0 | y) =
1-(0.362-0.267)=0.905$, which agrees extremely well to the exact
$p({\calH}_0 | y) = 0.907$.  The spherical shell is approximated by
(\ref{eq:h03}) only locally in a neighborhood of $y$ but not as a
whole.  Nevertheless, Theorem~\ref{thm:twosided} worked fine.

We next think of the situation that bootstrap probabilities of
${\calH}_1$ and ${\calH}_2$ are available but not their exact
$p$-values.  We apply the procedure of Section~\ref{sec:two} separately
to the two regions for calculating the approximately unbiased
$p$-values. To work on the procedure, here we consider a simple model
$\psi(\sigma^2)=\beta_0 + \beta_1 \sigma^2$ with parameters
$\varphi=(\beta_0, \beta_1)$ for
\begin{equation} \label{eq:fh1}
 \alpha_{\sigma^2}({\calH}_0'|y) = \Phi(-\psi(\sigma^2)/\sigma).
\end{equation}
Let $\psi_i(\sigma^2)$ be the normalized bootstrap $z$-value of
${\calH}_i$ for $i=1,2$. By assuming the simple model for
$\psi_i(\sigma^2)$, we fit $\alpha_{\sigma^2}({\calH}_i|y) =
\Phi(-\psi_i(\sigma^2)/\sigma)$ to the observed multiscale bootstrap
probabilities of ${\calH}_i$ for estimating the parameters.  The actual
estimation was done using the method described in
Section~\ref{sec:highjointbp}, but we would like to forget the details
for the moment.  We get $\hat\beta_0 = 0.101$, $\hat\beta_1 = -0.258$
for ${\calH}_1$, and similarly $\hat\beta_0 = 0.889$, $\hat\beta_1 =
0.286$ for ${\calH}_2$.  $\beta_0$'s are interpreted as the distances
from $y$ to the boundary surfaces, and the estimates agree well to the
exact values $\beta_0=0.1$ for ${\calH}_1$ and $\beta_0=0.9$ for
${\calH}_2$.  Then the approximately unbiased $p$-values are computed by
(\ref{eq:ph02}) as $p({\calH}_1|y)=\Phi(-0.101-0.258)=0.360$ and
$p({\calH}_2|y)=\Phi(-0.889+0.286)=0.273$, and thus (\ref{eq:ph03})
gives $p({\calH}_0 | y) = 1-(0.360-0.273)=0.913$, which again agrees
well to the exact $p({\calH}_0 | y) = 0.907$.

We finally think of a more practical situation, where the bootstrap
probabilities are not available for ${\calH}_1$ and ${\calH}_2$, but
only for ${\calH}_0$.  This situation is plausible in applications where
many regions are involved and we are not sure which of them can be
treated as ${\calH}_1$ or ${\calH}_2$ in a neighborhood of $y$; See
\cite{bib:Efro:Hall:Holm:96:BCL} for an illustration.  We consider a
simple model $\psi_1(\sigma^2)=\beta_0+\beta_1 \sigma^2$,
$\psi_2(\sigma^2)=d-\beta_0-\beta_1 \sigma^2$ with parameters
$\varphi=(\beta_0,\beta_1,d)$ for
\begin{equation} \label{eq:fh0}
 \alpha_{\sigma^2}({\calH}_0|y) = 1 - (
\Phi(-\psi_1(\sigma^2)/\sigma)+\Phi(-\psi_2(\sigma^2)/\sigma))
\end{equation}
by assuming that the two surfaces are curved in the same direction with
the same magnitude of curvature $|\beta_1|$.  For estimating $\varphi$,
(\ref{eq:fh0}) is fitted to the observed multiscale bootstrap
probabilities of ${\calH}_0$ with constraints $\beta_0>-d/2$ and $d>0$,
and $\hat\varphi$ is obtained as $\hat\beta_0=0.089$,
$\hat\beta_1=-0.199$, $\hat d = 0.995$.  Then the approximately unbiased
$p$-values are computed by (\ref{eq:ph02}) as $p({\calH}_1|y)=\Phi(-
0.089 - 0.199) = 0.387$ and $p({\calH}_2|y)=\Phi(- 0.995 + 0.089 +
0.199) = 0.240$ and thus (\ref{eq:ph03}) gives $p({\calH}_0 | y) =
1-(0.387 - 0.240) = 0.853$.  This is not very close to the exact
$p({\calH}_0 | y) = 0.907$, partly because the model is too simple.
However, it is a great improvement over $\alpha_1({\calH}_0 | y) =
P(a_1^2 \le \chi_{m+1}^2(\|y\|^2) \le a_2^2) = 0.320$.


\section{Bayesian measures of confidence} \label{sec:bayes}

Choosing a good prior density is essential for Bayesian inference.  We
consider a version of noninformative prior for making the posterior
probability acquire frequentist properties.

First note that the sum of bootstrap probabilities of disjoint
partitions of the whole parameter space is always 1.  For the two
regions case, $\alpha_{\sigma^2}({\calH}_0'|y) +
\alpha_{\sigma^2}({\calH}_1|y) = 1$, and thus $ \sigma
z_{\sigma^2}({\calH}_0'|y) + \sigma z_{\sigma^2}({\calH}_1|y) = 0$.
Therefore $p({\calH}_0'|y) + p({\calH}_1|y) = 1$ for the approximately
unbiased $p$-values computed by (\ref{eq:ph02}), suggesting that we may
think of a prior so that $p({\calH}_0'|y)=\pi({\calH}_0'|y)$.  This was
the idea of \cite{bib:Efro:Tibs:98:PR} to define a Bayesian measure of
confidence of ${\calH}_0$.
Since each of ${\calH}_1$ and ${\calH}_2$ can be treated as ${\calH}_0'$ by changing the coordinates, we may assume a prior satisfying
\begin{equation} \label{eq:pmatch2}
 \pi({\calH}_i|y) =  p({\calH}_i|y),\quad i=1,2.
\end{equation}
It follows from
 $\sum_{i=0}^2 \pi({\calH}_i|y)=1$ that
\begin{equation} \label{eq:pp03}
\pi({\calH}_0|y) = 1 - ( p({\calH}_1|y) +
p({\calH}_2 | y) ).
\end{equation}

Priors satisfying (\ref{eq:pmatch2}) are called probability matching
priors. The theory has been developed in literature
\citep{bib:Peers:1965:CPB,bib:Tibshirani:1989:NPO,bib:Datta:Mukerjee:2004:PMP}
for posterior quantiles of a single parameter of interest.  The examples
are the flat prior $\pi(\mu)=$ const for the flat boundary case in
Section~\ref{sec:intro}, and $\pi(\mu)\propto \|\mu\|^{-m}$ for the
spherical shell case in Section~\ref{sec:three}.

Our multiscale bootstrap method provides a new computation to
$\pi({\calH}_0|y)$.  We may simply compute (\ref{eq:pp03}) with the
$p({\calH}_1|y)$ and $p({\calH}_2|y)$ used for computing
$p({\calH}_0|y)$ of (\ref{eq:ph03}).  Although we implicitly assumed the
matching prior, we do not have to know the functional form of
$\pi(\mu)$.  For the spherical shell example, we may use the exact
$p({\calH}_1|y)$ and $p({\calH}_2|y)$ to get $p({\calH}_0 | y) =
1-(0.362+0.267)=0.371$, or more practically, use only bootstrap
probabilities of ${\calH}_0$ to get $p({\calH}_0 | y) = 1-(0.387 +
0.240) = 0.373$.

\section{Estimating parametric models for the scaling-law of bootstrap probabilities} \label{sec:model}

\subsection{One-step multiscale bootstrap} \label{sec:onestep}

We first recall the estimation procedure of
\cite{bib:Shim:2002:AUT,bib:Shimo:2008:TRN} before describing our new
proposals for improving the estimation accuracy in the following
sections.

Let $f(\sigma^2|\varphi)$ be a parametric model of bootstrap probability
such as (\ref{eq:fh1}) for ${\calH}_0'$ or (\ref{eq:fh0}) for
${\calH}_0$. As already mentioned in Section~\ref{sec:intro}, the model
is fitted to the observed $C_i/B_i$, $i=1,\ldots, M$. Since $C_i$ is
distributed as binomial with probability $f(\sigma_i^2|\varphi)$ and
$B_i$ trials, the log-likelihood function is $
\ell(\varphi)=\sum_{i=1}^M \{ C_i \log f(\sigma_i^2|\varphi) + (B_i-C_i)
\log (1-f(\sigma_i^2|\varphi)) \}$.  The maximum likelihood estimate
$\hat\varphi$ is computed numerically for each model. Let $\dim \varphi$
denote the number of parameters. Then $AIC = -2\ell(\hat\varphi)+2 \dim
\varphi$ may be compared for selecting a best model among several
candidate models.

\subsection{Two-step multiscale bootstrap} \label{sec:twostep}

\cite{bib:Shim:2004:AUT} has devised the multistep-multiscale bootstrap
as a generalization of the multiscale bootstrap. The usual multiscale
bootstrap is a special case called as the one-step multiscale bootstrap.
Our new proposal here is to utilize the two-step multiscale bootstrap
for improving the estimation accuracy of $\varphi$, although the
two-step method was originally used for replacing the normal model of
(\ref{eq:ynorm}) with the exponential family of distributions.

Recalling that ${\cal X}^*$ is obtained by resampling from ${\cal X}$,
we may resample again from ${\cal X}^*$, instead of ${\cal X}$, to get a
bootstrap sample of size $n''$, and denote it as ${\cal
X}^{**}=\{x^{**}_1,\ldots,x^{**}_{n''}\}$.  We formally consider the
parametric bootstrap
\[
 Y^{**}|y^* \sim
N_{m+1}(y^*,(\tau^2-\sigma^2)I_{m+1}),
\]
where $\tau$ is a new scale defined by $\tau^2-\sigma^2=n/n''$. In
\cite{bib:Shim:2004:AUT}, only the marginal distribution $ Y^{**}|y \sim
N_{m+1}(y,\tau^2 I_{m+1})$ is considered to detect the nonnormality. For
the second step, $ P_{\sigma^2,\tau^2}(Y^{**} \in {\calH}_0|y) =
\alpha_{\tau^2}({\calH}_0|y) $ should have the same functional form as
$P_{\sigma^2,\tau^2}(Y^* \in {\calH}_0|y) =
\alpha_{\sigma^2}({\calH}_0|y)$ for the normal model.  Here we also
consider the joint distribution of $(Y^*,Y^{**})$ given $y$. It is
$2m+2$-dimensional multivariate normal with $Cov(Y^*,Y^{**}|y) =
\sigma^2 I_{m+1}$. We denote the probability and the expectation by
$P_{\sigma^2,\tau^2}(\cdot|y)$ and $E_{\sigma^2,\tau^2}(\cdot|y)$,
respectively. Then, the joint bootstrap probability is defined as
\[
 \alpha_{\sigma^2,\tau^2}({\calH}_0|y)= P_{\sigma^2,\tau^2}(Y^* \in {\calH}_0 \wedge Y^{**}
\in {\calH}_0|y).
\]

Let $g(\sigma^2,\tau^2|\varphi)$ be a parametric model of
$\alpha_{\sigma^2,\tau^2}({\calH}_0'|y)$ or
$\alpha_{\sigma^2,\tau^2}({\calH}_0|y)$. To work on specific forms of
$g(\sigma^2,\tau^2|\varphi)$, we need some notations.  Let $(X',X'')$ be
distributed as bivariate normal with mean $(0,0)$, variance
$V(X')=V(X'')=1$, and covariance $Cov(X',X'')=\rho$.  The distribution
function is denoted as $\Phi_\rho(a_1,b_1)=P(X'\le a_1 \wedge X''\le
b_1)$, where the joint density is explicitly given as
$\phi_\rho(a_1,b_1) = (1- \rho^2)^{-1/2} \phi((1-\rho^2)^{-1/2} (b_1 -
\rho a_1) )\phi(a_1)$.  We also define $\Phi_\rho(a_1,b_1; a_2,b_2) =
P(a_2\le X' \le a_1 \wedge b_2\le X'' \le b_1) =
\Phi_\rho(a_1,b_1)-\Phi_\rho(a_2,b_1) - \Phi_\rho(a_1,b_2) +
\Phi(a_2,b_2)$. Then a generalization of (\ref{eq:scalelawh02}) is given
as follows. The proof is in Appendix~\ref{app:proof-jointbp}.
\begin{lemma}\label{lem:jointbp}
For sufficiently small $\lambda$, the joint bootstrap probabilities for
 ${\calH}_0'$ and ${\calH}_0$ are expressed asymptotically as
\begin{eqnarray}
  \alpha_{\sigma^2,\tau^2}({\calH}'_0|y) &= &
\Phi_\rho (z_1,w_1) + O(\lambda^2),\label{eq:gh1}\\
 \alpha_{\sigma^2,\tau^2}({\calH}_0|y) &= &
\Phi_\rho (z_1,w_1; z_2,w_2) + O(\lambda^2), \label{eq:gh0}
\end{eqnarray}
where $z_1 = -(v+{\cal E}_{\sigma^2} h_1(u))/\sigma$, 
$w_1 = -(v+{\cal E}_{\tau^2} h_1(u))/\tau$,
$z_2 = -(v+d+{\cal E}_{\sigma^2} h_2(u))/\sigma$, 
$w_2 = -(v+d+{\cal E}_{\tau^2} h_2(u))/\tau$, and
$\rho=\sigma/\tau$.
\end{lemma}
Thus $g(\sigma^2,\tau^2|\varphi)$ is specified for ${\calH}_0'$ as
(\ref{eq:gh1}) with $z_1 = -\psi(\sigma^2)/\sigma$, $w_1 =
-\psi(\tau^2)/\tau$ using the $\psi$ function of
(\ref{eq:fh1}). Similarly, $g(\sigma^2,\tau^2|\varphi)$ is specified for
${\calH}_0$ as (\ref{eq:gh0}) with $z_1 = \psi_1(\sigma^2)/\sigma$,
$w_1 = \psi_1(\tau^2)/\tau$, $z_2 = -\psi_2(\sigma^2)/\sigma$, $w_2 =
-\psi_2(\tau^2)/\tau$ using $\psi_1$ and $\psi_2$ functions of
(\ref{eq:fh0}).

We may specify $M$ sets of $(\sigma,\tau)$, denoted as
$(\sigma_1,\tau_1),\ldots, (\sigma_M,\tau_M)$. In our numerical
examples, $\sigma_1,\ldots,\sigma_{13}$ are specified as mentioned in
Section~\ref{sec:intro} and $\tau_i$'s are specified so that
$\tau_i^2-\sigma_i^2 = 1$ holds always, meaning $n''=n$. For each
$i=1,\ldots, M$, we generate $(Y^*,Y^{**})$ with $(\sigma_i,\tau_i)$
many times, say $B_i=10000$, and observe the frequencies $C_i = \#(Y^*
\in {\calH}_0)$, $D_i = \#(Y^{**} \in {\calH}_0)$, and $E_i = \#(Y^* \in
{\calH}_0 \wedge Y^{**} \in {\calH}_0)$.  Note that only one $Y^{**}$ is
generated from each $Y^*$ here, whereas thousands of $Y^{**}$'s may be
generated from each $Y^*$ in the double bootstrap method.  The
log-likelihood function becomes $\ell(\varphi)= \sum_{i=1}^M \{ E_i \log
g(\sigma^2_i,\tau^2_i|\varphi) + (C_i-E_i) \log(f(\sigma^2_i|\varphi) -
g(\sigma^2_i,\tau^2_i|\varphi) )+ (D_i-E_i) \log(f(\tau^2_i|\varphi) -
g(\sigma^2_i,\tau^2_i|\varphi) ) + (B_i -C_i -D_i+E_i) \log(1-
f(\sigma^2_i|\varphi)-f(\tau^2_i|\varphi) +
g(\sigma^2_i,\tau^2_i|\varphi) ) \}$. In fact, we have used this
two-step multiscale bootstrap, instead of the one-step method, in all
the numerical examples.

The one-step method had difficulty in distinguishing ${\calH}_0$ with
very small $d$ from ${\calH}_0$ with moderate $d$ but heavily curved
$\partial {\calH}_1$. The two-step method avoids this identifiability
issue because a small value of $E_i$ indicates that $d$ is small; It is
automatically done, of course, by the numerical optimization of
$\ell(\varphi)$.

\subsection{Higher-order terms of bootstrap probabilities for
testing two regions} \label{sec:highjointbp}

The asymptotic errors of the scaling law of the bootstrap probabilities
in (\ref{eq:bph02}) and (\ref{eq:gh1}) are of order $O(\lambda^2)$.  As
shown in the following lemma, the errors can be reduced to
$O(\lambda^3)$ by introducing correction terms of $O(\lambda^2)$ for
improving the parametric model $g(\sigma^2,\tau^2|\varphi)$ of
${\calH}_0'$.  The proof is given in Appendix~\ref{app:proof-jointbph}.
\begin{lemma}\label{lem:jointbph}
For sufficiently small $\lambda$, the bootstrap probabilities for
${\calH}_0'$ are expressed asymptotically as
\begin{eqnarray}
 \alpha_{\sigma^2}({\calH}_0'|y) &=& \Phi(z_1+\Delta z_1) +
  O(\lambda^3)
\label{eq:ach1} \\
 \alpha_{\tau^2}({\calH}_0'|y) &=& \Phi(w_1+\Delta w_1) +
  O(\lambda^3)
\label{eq:ach2} \\
 \alpha_{\sigma^2,\tau^2}({\calH}'_0 |y) &=& 
\Phi_{\rho+\Delta \rho} (z_1+\Delta z_1,w_1+\Delta w_1) + O(\lambda^3),
\label{eq:achjoint}
\end{eqnarray}
where $z_1$, $w_1$, and $\rho$ are those defined in Lemma~\ref{lem:jointbp},
 and the higher order correction terms are defined as $ \Delta z_1 =
 -\frac{1}{2} z_1 E_{\sigma^2,\tau^2}(\epsilon_1^2 | u)$, $ \Delta w_1 =
 -\frac{1}{2} w_1 E_{\sigma^2,\tau^2}(\delta_1^2 | u) $, and $
 \Delta\rho=-\frac{1}{2}\left( \rho E_{\sigma^2,\tau^2}(\epsilon_1^2 |
 u) + \rho E_{\sigma^2,\tau^2}(\delta_1^2 | u) - 2
 E_{\sigma^2,\tau^2}(\epsilon_1 \delta_1 | u) \right)$ using
\begin{equation} \label{eq:e1e2}
 \epsilon_1 = -\frac{h_1(U^*) - {\cal E}_{\sigma^2} h_1(u)}{\sigma},\quad
 \delta_1 = -\frac{h_1(U^{**}) - {\cal E}_{\tau^2} h_1(u)}{\tau}.
\end{equation}
\end{lemma}

For deriving a very simple model for $\Delta \rho$, we think of a
situation $h(u) = (A/\!\sqrt{m}) \|u\| + (B/m) \|u\|^2$ and $\theta=0$,
and consider asymptotics as $m\to\infty$. This formulation is only for
convenience of derivation.  The two values $A$ and $B$ will be specified
later by looking at the functional form of $f(\sigma^2|\varphi)$.  A
straightforward, yet tedious, calculation (the details are not shown)
gives $\psi(\sigma^2)=\mbox{const}+A\sigma +B\sigma^2+O(m^{-1})$ and
\[
 \Delta\rho = -\frac{1}{2m}\left(
A^2 \rho(1-\rho) + 2B^2 \rho(\tau^2-\sigma^2)
+ 2 A B \sigma (1-\rho^2)
\right) + O(m^{-3/2}).
\]

This correction term was in fact already used for the simple model
$\psi(\sigma^2)=\beta_0 + \beta_1 \sigma^2$ of the spherical shell
example in Section~\ref{sec:three}, where the parameter was actually
$\varphi=(\beta_0,\beta_1,m)$ instead of $\varphi=(\beta_0,\beta_1)$. We
did not change the $\psi(\sigma^2)$ for adjusting $\Delta z_1$ and
$\Delta w_1$, meaning that $z_1 + \Delta z_1$, instead of $z_1$, was
modelled as $-\psi(\sigma^2)/\sigma$.  Comparing the coefficients of
$\psi(\sigma^2)$, we get $A=0$ and $B=\beta_1$, and thus $\Delta \rho= -
(\beta_1)^2 (\sigma/\tau) (\tau^2-\sigma^2)/m$. When (\ref{eq:fh1}) was
fitted to ${\calH}_1$, the estimated parameter $\hat m=2.83$ was close
to the true value $m=3$.


For the numerical example mentioned above, we have also fitted the same
model but $\Delta \rho=0$ being fixed. The estimated parameters are
$\hat\beta_0=0.101$, $\hat\beta_1=-0.256$, and the $p$-value is $p({\calH}_1|y)=\Phi(-0.101-0.256)=0.361$. These values are not much different
from those shown in Section~\ref{sec:three}. However, the AIC value
improved greatly by the introduction of $\Delta \rho$, and the AIC
difference was 96.67, mostly because improved fitting for the joint
bootstrap probability of (\ref{eq:achjoint}).  My experience suggests
that consideration of the $\Delta \rho$ term is useful for choosing a
reasonable model of $\psi(\sigma^2)$.


\section{Simulation study} \label{sec:simulation}

Let us consider a cone-shaped region ${\calH}_0$ in $\mathbb{R}^2$ with
the angle at the vertex being $2\pi/10$ as shown in Panel~(b) of
Fig.~\ref{fig:exregions}. This cone can be regarded, locally in a
neighborhood of $y$ with appropriate coordinates, as ${\calH}_0$ of
(\ref{eq:h03}) when $y$ is close to one of the edges but far from the
vertex, or as ${\calH}_0'$ of (\ref{eq:h02}) when $y$ is close to the
vertex. In this section, the cone is labelled either by ${\calH}_0$ or
${\calH}_0'$ depending on which view we are taking.

Cones in $\mathbb{R}^2$ appear in the problem of multiple comparisons of
three elements $X_0, X_1, X_2$, say, and ${\calH}_i$ corresponds to the
hypothesis that the mean of $X_i$ is the largest among the three
\citep{bib:Dupr:Swan:Vent:Some:85:SPS,bib:Perl:Wu:2003:OVL,bib:Shimo:2008:TRN}.
The angle at the vertex is related to the covariance structure of the
elements.  Although an unbiased test does not exist for this region, we
would like to see how our methods work for reducing the coverage error.

\begin{figure}
\includegraphics[width=0.70\textwidth]{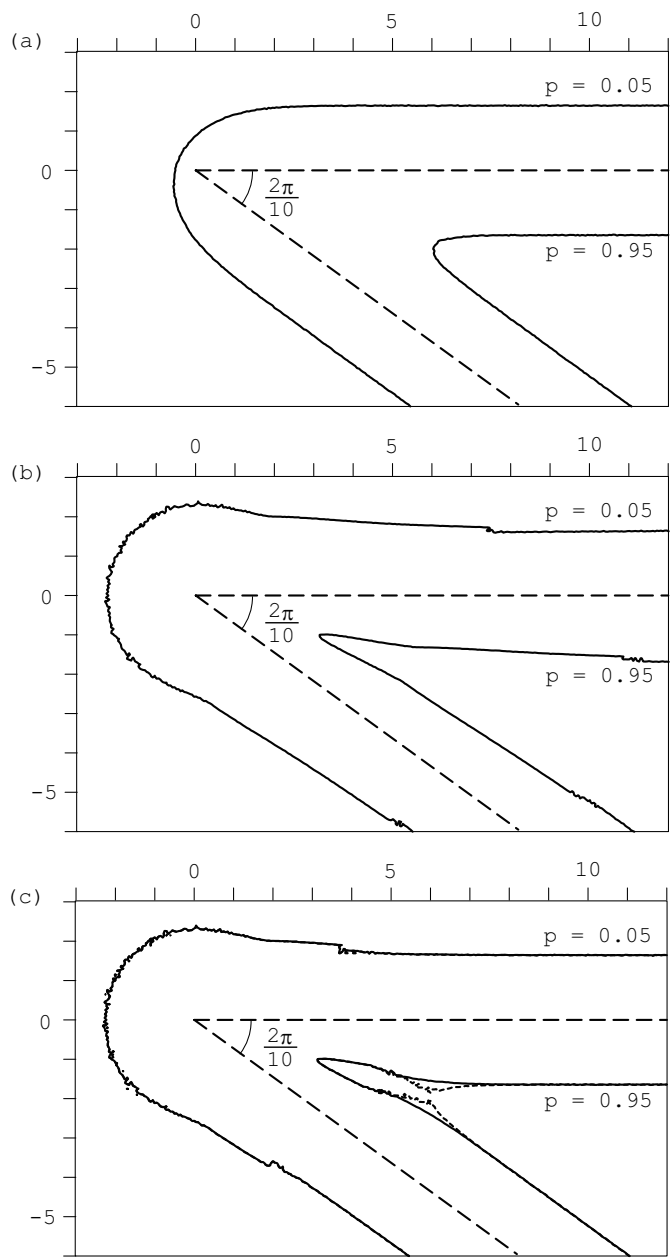}
\caption{Contour lines $p(y)=0.05$ and $p(y)=0.95$. The cone-shaped
region ${\calH}_0$ is rotated so that one of the edges is placed
along the x-axis.  Solid curves are drawn for (a)~the bootstrap
probability with $\sigma^2=1$, and for (b)~the frequentist $p$-value for
``one-sided'' test.  In Panel~(c), $p(y)$ is switched to the frequentist
$p$-value for ``two-sided'' test when appropriate.  The dotted curve in
Panel~(c) is for the Bayesian posterior probability.}
\label{fig:rejregion10}
\end{figure}

\begin{figure}
\includegraphics[width=0.75\textwidth]{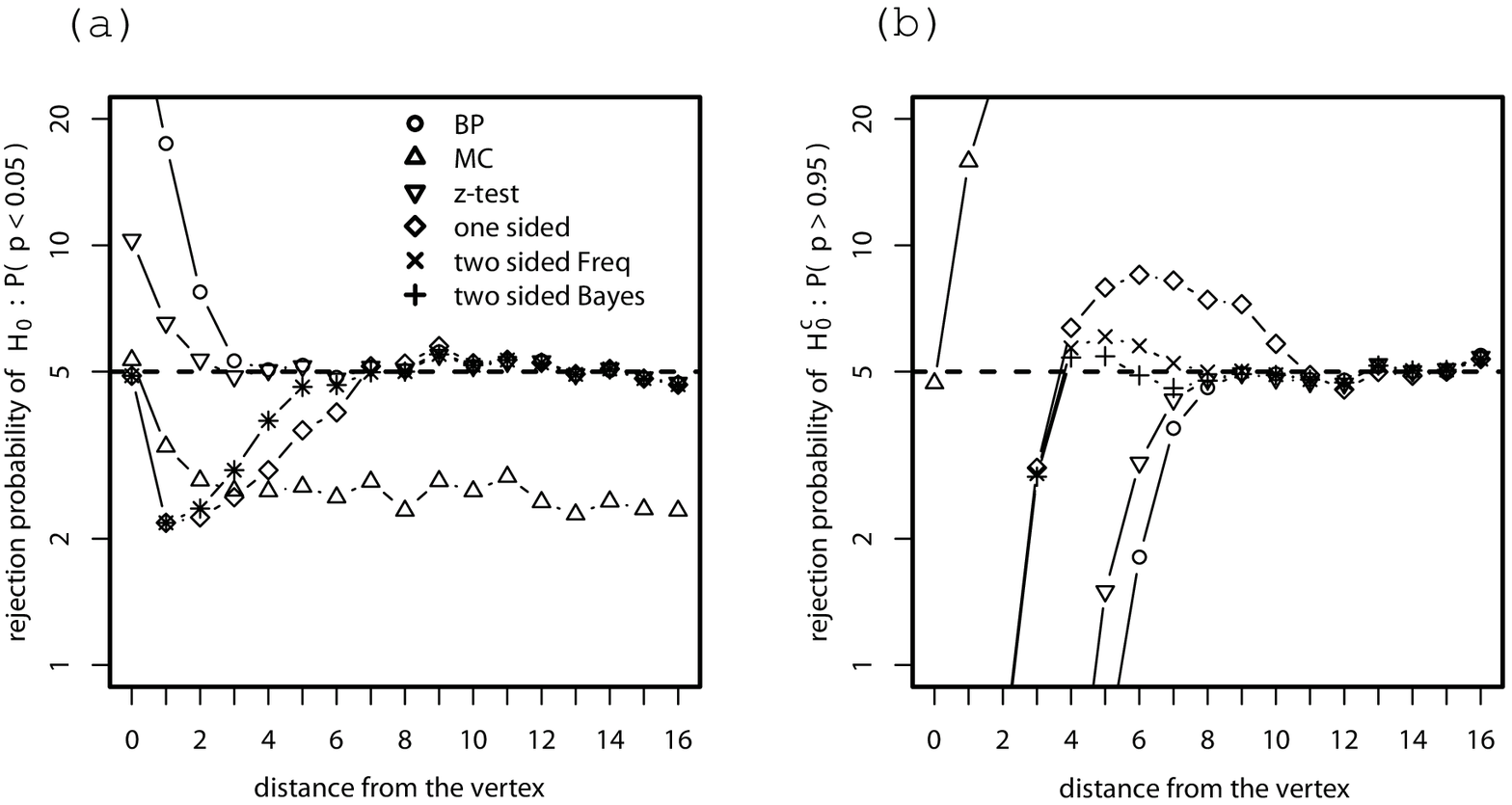}
\caption{(a)~Rejection probability of the cone, and (b)~that of the
 complement of the cone. The angle at the vertex is $2\pi/10$.}
\label{fig:rejprob10}  
\end{figure}

\begin{figure}
\includegraphics[width=0.75\textwidth]{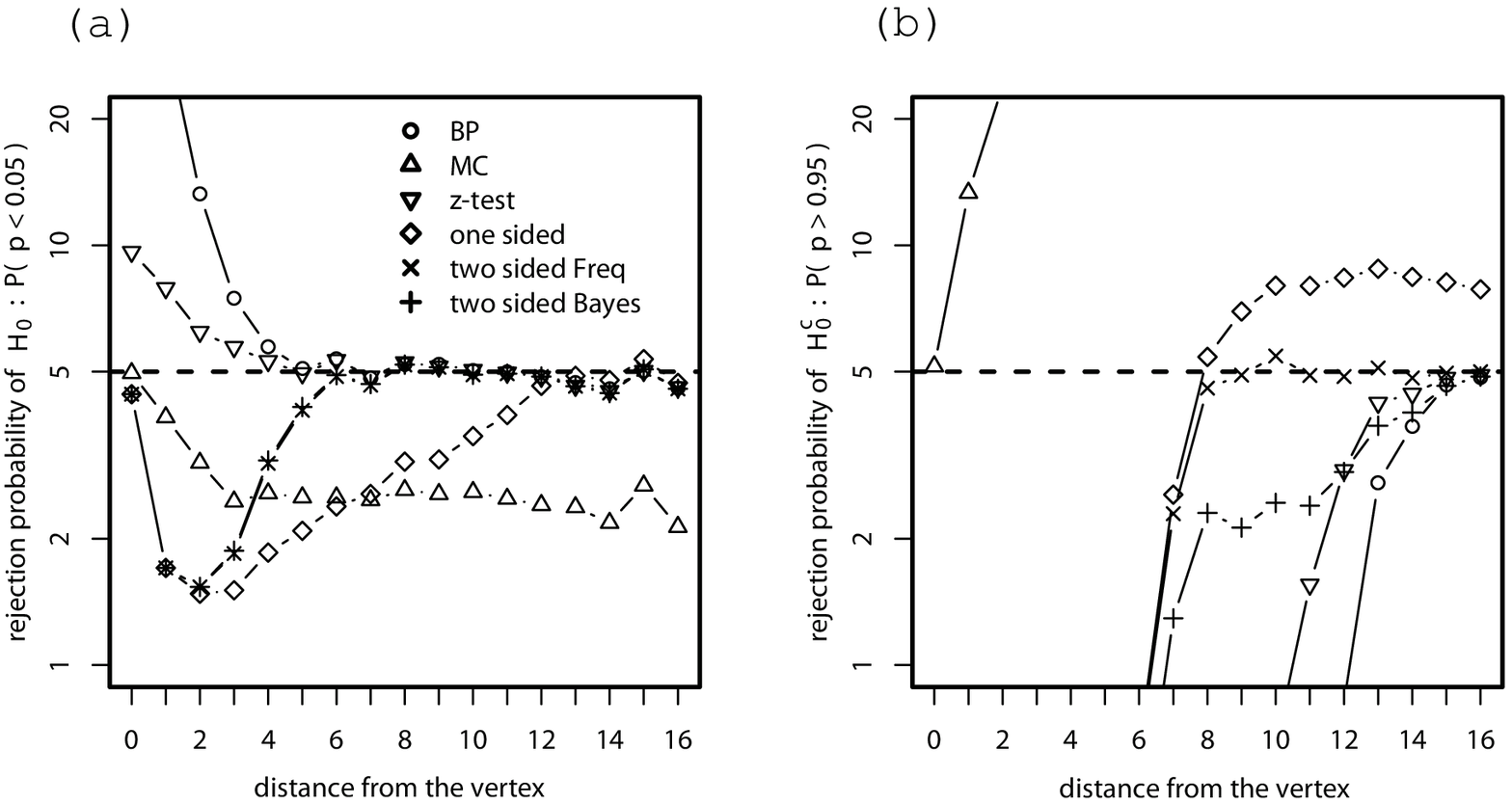}
\caption{(a)~Rejection probability of the cone, and (b)~that of the
 complement of the cone. The angle at the vertex is $2\pi/20$.}
\label{fig:rejprob20}  
\end{figure}

Contour lines of confidence measures, denoted $p(y)$ in general, at the
levels 0.05 and 0.95 are drawn in Fig.~\ref{fig:rejregion10}.  The
rejection regions of the cone and the complement of the cone are
${\calR}=\{y | p(y)<0.05\}$ and ${\calR}'=\{y | p(y)>0.95\}$,
respectively, at $\alpha=0.05$.  We observe that $p(y)$ decreases as $y$
moves away from the cone in Panels~(a), (b), and (c); See
Appendix~\ref{app:sim} for the details of computation. On the other hand,
Figs.~\ref{fig:rejprob10} and \ref{fig:rejprob20} show the rejection
probability. For an unbiased test, it should be 5\% for all the
$\mu\in\partial{\calH}_0$ so that the coverage error is zero.

In Panel~(a) of Fig.~\ref{fig:rejregion10}, $p(y)=\alpha_1({\calH}_0|y)$
is computed by the bootstrap samples of $\sigma^2=1$. This bootstrap
probability, labelled as BP in Fig.~\ref{fig:rejprob10}, is heavily
biased near the vertex, and this tendency is enhanced when the angle
becomes $2\pi/20$ in Fig.~\ref{fig:rejprob20}.

In Panel~(b) of Fig.~\ref{fig:rejregion10}, $p(y)=p({\calH}_0'|y)$ is
computed by regarding the cone as ${\calH}_0'$ of (\ref{eq:h02}).  The
dent of ${\calR}$ and the bump of ${\calR}'$ become larger than those of
Panel~(a) of Fig.~\ref{fig:rejregion10} near the vertex, confirming what
we observed in \cite{bib:Shimo:2008:TRN}. As seen in
Figs.~\ref{fig:rejprob10} and \ref{fig:rejprob20}, the coverage error of
$p({\calH}_0'|y)$, labelled as ``one sided'' there, is smaller than that
of BP.

In Panel~(c) of Fig.~\ref{fig:rejregion10}, $p({\calH}_0|y)$ is also
computed by regarding the cone as ${\calH}_0$ of (\ref{eq:h03}), and
then one of $p({\calH}_0'|y)$ and $p({\calH}_0|y)$ is selected as $p(y)$
by comparing the AIC values at each $y$. This $p(y)$, labelled as ``two
sided Freq'' in Figs.~\ref{fig:rejprob10} and \ref{fig:rejprob20},
improves greatly on the one-sided $p$-value.  The coverage error is
almost zero except for small $\|\mu\|$'s, verifying what we attempted in
this paper.  The corresponding Bayesian posterior probability, labelled
as ``two sided Bayes,'' performs similarly.  Note that the coverage
error was further reduced near the vertex by setting simply
$p(y)=p({\calH}_0|y)$ without the model selection (the result is not
shown here); However, the shapes of $\calR$ and ${\calR}'$ became rather
weird then in the sense mentioned at the last paragraph of
Section~\ref{sec:two}.

\section{Concluding Remarks} \label{sec:discussion}

In this paper, we have discussed frequentist and Bayesian measures of
confidence for the three regions case, and have proposed a new
computation method using the multiscale bootstrap technique. In this
method, AIC played an important role for choosing appropriate parametric
models of the scaling-law of bootstrap probability.  Simulation study
showed that the proposed frequentist measure performs better for
controlling the coverage error than the previously proposed multiscale
bootstrap designed only for the two regions case.

A generalization of the confidence measures gives a frequentist
interpretation of the Bayesian posterior probability as follows.  Let us
consider the situation of Theorem~\ref{thm:twosided}.  If we strongly
believe that $\mu \not \in {\calH_2}$, we could use the one-sided
$p$-value $p({\calH}_0'|y)=1-p({\calH}_1|y)$, instead of the two sided
$p({\calH}_0|y)$. Similarly, we might use $1-p({\calH}_2|y)$ if we
believe that $\mu\not\in {\calH}_1$.  By making the choice
``adaptively,'' someone may want to use $ p^{(1)} ({\calH}_0|y) = 1
-\max(p({\calH}_1|y) , p({\calH}_2|y))$, although it is not justified in
terms of coverage error. By connecting $p^{(1)} ({\calH}_0|y)$ and
$p({\calH}_0|y)$ linearly using an index $s$ for the number of
``sides,'' we get
\[
 p^{(s)} ({\calH}_0|y) = \pi({\calH}_0|y) + s \min(p({\calH}_1|y) ,
 p({\calH}_2|y)).
\]
It is easily verified that $p({\calH}_0|y)=p^{(2)} ({\calH}_0|y)$ and
$\pi({\calH}_0|y)=p^{(0)} ({\calH}_0|y)$, indicating that the Bayesian
posterior probability defined in Section~\ref{sec:bayes} can be
interpreted, interestingly, as a frequentist $p$-value of ``zero-sided''
test of ${\calH}_0$. Although we have no further consideration, this
kind of argument might lead to yet another compromise between
frequentist and Bayesian.

Our formulation is rather restrictive. We have considered only the three
regions case by introducing the surface $h_2$ in addition to the surface
$h_1$ of the two regions case. Also these two surfaces are assumed to be
nearly parallel to each other. It is worth to elaborate on
generalizations of this formulation in future work, but too much of
complication may result in unstable computation for estimating the
scaling-law of bootstrap probability. AIC will be useful again in such a
situation.

\appendix
\section{Proofs}


\subsection{Proof of Theorem~\ref{thm:twosided}} \label{app:proof-twosided}

First we consider rejection regions of testing ${\calH}_0$ for a
specified $\alpha$ by modifying the two rejection regions of
(\ref{eq:h0s3}). Since $h_1$ and $h_2$ are nearly flat, the modified
regions should be expressed as ${\calR}_1 = \{(u,v) \mid v>c -
r_1(u),\,u\in\mathbb{R}^m\}$ and ${\calR}_2 = \{(u,v) \mid v<-d-c -
r_2(u),\,u\in\mathbb{R}^m\}$ using nearly flat functions $r_1$ and
$r_2$. The constant $c$ is the same one as defined in (\ref{eq:defc}).
Write $a=\phi(c)$, $b=\phi(c+d)$ for brevity sake. We evaluate the
rejection probability for $\mu\in\partial {\calH}_1 \cup \partial
{\calH}_2$. Let $\mu\in\partial {\calH}_1$ for a moment, and put $\mu =
(\theta,-h_1(\theta))$. By applying the argument of (\ref{eq:bph02}) to
${\calR}_1$ but (\ref{eq:yboot}) is replaced by (\ref{eq:ynorm}), we get
$P(Y\in{\calR}_1|\mu)=1-\Phi(c - {\cal E}_1 r_1(\theta) + h_1(\theta)) +
O(\lambda^2)=\Phi(-c) + a({\cal E}_1 r_1(\theta) -
h_1(\theta))+O(\lambda^2)$. The same argument applied to ${\calR}_2$
gives $P(Y\in{\calR}_2|\mu)=\Phi(-d-c-{\cal E}_1
r_2(\theta)+h_1(\theta))+O(\lambda^2) =\Phi(-d-c) + b(-{\cal E}_1
r_2(\theta)+h_1(\theta))+O(\lambda^2)$. Rearranging these two formula
with the identity
\begin{equation} \label{eq:prftwo-a}
 P(Y\in{\calR}_1|\mu) + P(Y\in{\calR}_2|\mu)=\alpha
\end{equation}
for an unbiased test, we get an equation $ a ({\cal
E}_1 r_1(\theta)-h_1(\theta))+ b(-{\cal E}_1 r_2(\theta)+h_1(\theta)) =
O(\lambda^2)$.  By exchanging the roles of $r_1$ and $r_2$, the equation
becomes $ b ({\cal E}_1 r_1(\theta)-h_2(\theta))+ a(- {\cal E}_1
r_2(\theta)+h_2(\theta)) = O(\lambda^2)$ for $\mu\in\partial {\calH}_2$ with $\mu=(\theta,-d-h_2(\theta))$. These two equations are
expressed as
\begin{equation}
 \left(\begin{array}{cc}
a & -b \\
-b & a
\end{array}\right)
 \left(\begin{array}{c}
{\cal E}_1 r_1(\theta)\\ {\cal E}_1 r_2(\theta)
\end{array}\right)
=(a-b)
 \left(\begin{array}{c}
h_1(\theta)\\ h_2(\theta)
\end{array}\right)
+O(\lambda^2). \label{eq:prftwo-e}
\end{equation}
For solving this equation with respect to $r_1$ and $r_2$, first apply
the inverse matrix of the $2\times 2$ matrix from the left in
(\ref{eq:prftwo-e}), and then apply the inverse operator of ${\cal E}_1$
so that
\begin{equation}
 \left(\begin{array}{c}
r_1(u)\\ r_2(u)
\end{array}\right)
=\frac{1}{a+b}
 \left(\begin{array}{cc}
a & b \\
b & a
\end{array}\right)
 \left(\begin{array}{c}
{\cal E}_{-1} h_1(u)\\
{\cal E}_{-1} h_2(u)
\end{array}\right)
+O(\lambda^2). \label{eq:prftwo-r}
\end{equation}

Next we obtain an expression of $p$-value corresponding to the rejection
regions. $p({\calH}_0|y)$ is defined as the value of $\alpha$ for
which either of $y\in\partial {\calR}_1$ and $y\in\partial {\calR}_2$ holds. Note that $r_1$, $r_2$, and $c$ depend on $\alpha$. Let us
assume $y\in\partial {\calR}_1$ and thus $c=v+r_1(u)$ for a moment.
Write $a'=\phi(v) = a + O(\lambda)$, $b'=\phi(v+d) = b+O(\lambda)$ for
brevity sake. Recalling (\ref{eq:defc}), $p({\calH}_0|y) = \Phi(-c)+
\Phi(-d-c) = \Phi(-v-r_1(u)) + \Phi(-d-v-r_1(u)) = \Phi(-v) + \Phi(-d-v)
- (a'+b')r_1(u) + O(\lambda^2)$, where $r_1(u)$ in (\ref{eq:prftwo-r}) can
be expressed as
\[
 r_1(u) = \frac{a'}{a'+b'} {\cal E}_{-1}h_1(u) + 
\frac{b'}{a'+b'} {\cal E}_{-1}h_2(u) + O(\lambda^2).
\]
Therefore, $p({\calH}_0|y) = \Phi(-v)+ \Phi(-d-v) -a' {\cal E}_{-1}h_1(u)
-b' {\cal E}_{-1}h_2(u) +O(\lambda^2) = \Phi(-v - {\cal E}_{-1}h_1(u)) +
\Phi(-d-v - {\cal E}_{-1}h_2(u))+O(\lambda^2)$. By applying
(\ref{eq:phinvh02}) to ${\calH}_1$ and ${\calH}_2$, respectively,
we get $p({\calH}_1|y)=\Phi(v+{\cal E}_{-1} h_1(u))+O(\lambda^2)$ and $p({\calH}_2|y)=\Phi(-v-d-{\cal E}_{-1} h_2(u))+O(\lambda^2)$, and thus $p({\calH}_0|y)=1- p({\calH}_1|y) + p({\calH}_2|y) + O(\lambda^2)$. By
exchanging the roles of ${\calH}_1$ and ${\calH}_2$, we have
$p({\calH}_0|y)=1- p({\calH}_2|y) + p({\calH}_1|y) +
O(\lambda^2)$ for $y\in \partial {\calR}_2$. By taking the minimum of
these two expressions of $p({\calH}_0|y)$, we finally obtain
(\ref{eq:ph03}). This $p$-value satisfies (\ref{eq:prftwo-a}) with error
$O(\lambda^2)$, and thus (\ref{eq:eh03}) holds.

\subsection{Proof of Lemma~\ref{lem:jointbp}} \label{app:proof-jointbp}

The argument is very similar to (\ref{eq:bph02}) in the proof of
Theorem~\ref{thm:onesided}. Given $v, u^*, u^{**}$, the joint
distribution of $X' = (V^*-v)/\sigma$ and $X'' = (V^{**}-v)/\tau$ is
$\Phi_\rho$. Therefore, $P_{\sigma^2,\tau^2}( V^* \le -h_1(u^*) \wedge
V^{**} \le -h_1(u^{**}) |v,u^{*},u^{**}) = P_{\sigma^2,\tau^2}( X' \le
z_1+\epsilon_1 \wedge X'' \le w_1+\delta_1 |v,u^{*},u^{**})
=\Phi_\rho(z_1+\epsilon_1,w_1+\delta_1)$, where
$\epsilon_1$ and $\delta_1$ are defined in (\ref{eq:e1e2}).
Taking the expectation with respect to $(U^*,U^{**})$, we have
$\alpha_{\sigma^2,\tau^2}({\calH}'_0|y) = P_{\sigma^2,\tau^2}( V^* \le
-h_1(U^*) \wedge V^{**} \le -h_1(U^{**}) | y) =E_{\sigma^2,\tau^2}(
\Phi_\rho(z_1+\epsilon_1,w_1+\delta_1)|u)$. 
For proving (\ref{eq:gh1}), considering the Taylor series around
$(z_1,w_1)$,
we obtain
\begin{equation} \label{eq:expjointbp}
E_{\sigma^2,\tau^2}\left(
\Phi_\rho(z_1,w_1) +
\frac{\partial \Phi_\rho}{\partial z_1} \epsilon_1
+\frac{\partial \Phi_\rho}{\partial w_1} \delta_1
\big|u
\right) + O(\lambda^2)
\end{equation}
with
$E_{\sigma^2,\tau^2}(\epsilon_1|u)=E_{\sigma^2,\tau^2}(\delta_1|u)=0$
for completing the proof.

Next we show (\ref{eq:gh0}). The conditional probability given
$v,u^{*},u^{**}$ is $P_{\sigma^2,\tau^2}( -d-h_2(u^*)\le V^* \le
-h_1(u^*) \wedge -d-h_2(u^{**}) \le V^{**} \le -h_1(u^{**})
|v,u^{*},u^{**}) = P_{\sigma^2,\tau^2}( z_2+\epsilon_2 \le X' \le
z_1+\epsilon_1 \wedge w_2+\delta_2 \le X'' \le w_1+\delta_1
|v,u^{*},u^{**}) =
\Phi_\rho(z_1+\epsilon_1,w_1+\delta_1;z_2+\epsilon_2,w_2+\delta_2)$, where
\[
 \epsilon_2 = -\frac{h_2(U^*) - {\cal E}_{\sigma^2} h_2(u)}{\sigma},\quad
 \delta_2 = -\frac{h_2(U^{**}) - {\cal E}_{\tau^2} h_2(u)}{\tau}.
\]
Taking the expectation with respect to $(U^*,U^{**})$, we have
$ \alpha_{\sigma^2,\tau^2}({\calH}_0 |y) 
=P_{\sigma^2,\tau^2}( 
-d-h_2(U^*)\le
V^* \le -h_1(U^*) \wedge
-d-h_2(U^{**}) \le
V^{**} \le -h_1(U^{**}) | y)
=E_{\sigma^2,\tau^2}(
\Phi_\rho(z_1+\epsilon_1,w_1+\delta_1;
z_2+\epsilon_2,w_2+\delta_2)|u)$. We only have to consider the Taylor
series
\[
E_{\sigma^2,\tau^2}\left(
\Phi_\rho(z_1,w_1;z_2,w_2) +
\frac{\partial \Phi_\rho}{\partial z_1} \epsilon_1
+\frac{\partial \Phi_\rho}{\partial w_1} \delta_1
+\frac{\partial \Phi_\rho}{\partial z_2} \epsilon_2
+\frac{\partial \Phi_\rho}{\partial w_2} \delta_2
\big|u
\right) + O(\lambda^2)
\]
with
$E_{\sigma^2,\tau^2}(\epsilon_i|u)=E_{\sigma^2,\tau^2}(\delta_i|u)=0$,
$i=1,2$ for completing the proof.

\subsection{Proof of Lemma~\ref{lem:jointbph}} \label{app:proof-jointbph}

By considering a higher-order term of the Taylor series in
(\ref{eq:bph02}), we obtain $\alpha_{\sigma^2}({\calH}_0'|y)=E_{\sigma^2}(\Phi(z_1)+\phi(z_1)\epsilon_1 - \phi(z_1) z_1
\epsilon_1^2/2 | u) + O(\lambda^3) =\Phi(z_1) +\phi(z_1) \Delta z_1 +
O(\lambda^3) = \Phi(z_1 +\Delta z_1)+O(\lambda^3)$, proving
(\ref{eq:ach1}) as well as (\ref{eq:ach2}). On the other hand,
(\ref{eq:achjoint}) is shown by considering higher-order terms of the
Taylor series in (\ref{eq:expjointbp}) as
\begin{eqnarray*}
E_{\sigma^2,\tau^2}\Bigl(
\Phi_\rho(z_1,w_1) &+&
\frac{\partial \Phi_\rho}{\partial z_1} \epsilon_1
+\frac{\partial \Phi_\rho}{\partial w_1} \delta_1 \nonumber \\
&+&
\frac{1}{2}\Bigl(
\frac{\partial^2 \Phi_\rho}{\partial z_1^2}\epsilon_1^2
+2\frac{\partial^2 \Phi_\rho}{\partial z_1 \partial w_1}\epsilon_1 \delta_1+
\frac{\partial^2 \Phi_\rho}{\partial w_1^2}\delta_1^2
\Bigr)
\big|u
\Bigr) + O(\lambda^3).
\end{eqnarray*}
The proof completes by rearranging the above formula with
\[
 \frac{\partial^2 \Phi_\rho}{\partial z_1^2}=
-z_1  \frac{\partial \Phi_\rho}{\partial z_1} - \rho
\phi_\rho(z_1,w_1),\quad
 \frac{\partial^2 \Phi_\rho}{\partial z_1 \partial w_1}=
\phi_\rho(z_1,w_1),\quad
 \frac{\partial \Phi_\rho}{\partial \rho}=
\phi_\rho(z_1,w_1).
\]

\section{Simulation Details} \label{app:sim}

The contour lines in Fig.~\ref{fig:rejregion10} are drawn by computing
$p$-values at all grid points ($300\times 180$) of step size 0.05 in the
rectangle area; This huge computation was made possible by parallel
processing using up to 700 cpus. The computation takes a few minutes per
each grid point per cpu. Our algorithm is implemented as an experimental
version of the scaleboot package of \cite{bib:Shim:2006:sau}, which
will be included soon in the release version available from CRAN.

The rejection probabilities in Figs.~\ref{fig:rejprob10} and
\ref{fig:rejprob20} are computed by generating $y$ according to
(\ref{eq:ynorm}) for 10000 times, and then counting how many times
$p(y)<0.05$ or $p(y)>0.95$ is observed.  This computation is done for
each $\mu\in\partial {\calH}_0$ with the distance from the vertex
$\|\mu\|=0,1,\ldots,16$, i.e., $\mu=(0,0), (1,0), \ldots, (16,0)$ in the
coordinates of Fig.~\ref{fig:rejregion10}.

For computing $p({\calH}'_0|y)$ and $p({\calH}_0|y)$, the two-step
multiscale bootstrap described in Section~\ref{sec:twostep} was
performed with the $M=13$ sets of scales $(\sigma_i, \tau_i)$,
$i=1,\ldots, 13$, specified there. The parametric bootstrap,
instead of the resampling, was used for the simulation.  The number of
bootstrap samples has increased to $B_i=10^5$ for making the contour
lines smoother, while it was $B_i=10^4$ in the other results.

For $p({\calH}'_0|y)$, we have considered the singular model of
\cite{bib:Shimo:2008:TRN} defined as $\psi(\sigma^2)=\beta_0 + \beta_1 /
(1 + \beta_2 (\sigma - 1))$ for cones, and performed the model fitting
method described in Section~\ref{sec:highjointbp}.  From the Taylor
series of this $\psi(\sigma^2)$ around $\sigma=1$, we get $A=\beta_1
\beta_2 (3 - 2\beta_2) $, $B=\beta_1(\beta_2-1)^2$ for computing the
higher order correction term $\Delta \rho$. We have also considered
submodels by restricting some of $\varphi=(\beta_0,\beta_1,\beta_2,m)$
to specified values, and the minimum AIC model is chosen at each
$y$. The frequentist $p$-value is computed by (\ref{eq:pk}) with $k=3$
and $\sigma_0^2=1$.

For $p({\calH}_0|y)$, we have considered the same singular model for the
two surfaces by assuming they are curved in the opposite
directions with the same magnitude of curvature. More specifically, the
two $\psi$ functions in (\ref{eq:fh0}) are defined as
$\psi_1(\sigma^2)=\beta_0 + \beta_1 / (1 + \beta_2 (\sigma - 1))$ and
$\psi_2(\sigma^2)=d-\beta_0 + \beta_1 / (1 + \beta_2 (\sigma - 1))$. The
parameters $\varphi=(\beta_0,\beta_1,\beta_2,d)$ are estimated by the
model fitting method described in Section~\ref{sec:twostep}. Submodels
are also considered and model selection is performed using AIC. The
frequentist $p$-value is computed by (\ref{eq:ph03}), and the Bayesian
posterior probability is computed by (\ref{eq:pp03}).

The rejection probabilities of other two commonly used measures are
shown only for reference purposes; See \cite{bib:Shimo:2008:TRN} for the
details.  The rejection probability of the multiple comparisons, denoted
MC here, is always below 5\% in Panel~(a), and the coverage error
becomes zero at the vertex. On the other hand, the rejection probability
of the $z$-test is always below 5\% in Panel~(b), and the coverage error
reduces to zero as $\|\mu\|\to\infty$.


\bibliographystyle{spbasic}      

\end{document}